# Mixed Random Sampling of Frames method for counting number of motifs

M N Yudina[1], V N Zadorozhnyi[1], E B Yudin[2]

[1]Omsk state technical university, pr. Mira 11, 644050, Omsk, Russia
[2]Sobolev Institute of Mathematics, Siberian Branch of the Russian Academy of Sciences, 4 Acad. Koptyug avenue, 630090, Novosibirsk, Russia

E-mail: mg-and-all@mail.ru, zwn2015@yandex.ru, udinev@asoiu.com

**Abstract**. The problem of calculating the frequencies of network motifs on three and four nodes in large networks is considered. Telecommunications networks, cell molecular networks are investigated. The sizes of the investigated networks are hundreds of thousands of nodes and connections. These networks are represented in the form of directed and undirected simple graphs. Exact calculating requires huge computational resources for such large graphs. A method for calculating the frequencies of network motifs using the Monte Carlo method with control of an accuracy of calculations is proposed. The proposed effective method minimizes the value of the coefficient of variation.

## 1. Introduction

A conception of «network motifs» was proposed by Milo et al in Science [1]. They pointed out that large technical, social and molecular networks are natural to explore bottom-up. This means that you can try to identify the presence of identical small functional modules that occur more often than in a random (with randomized arcs) graph, and then explain their interaction in the network as a whole. Such small modules are called network motifs. In term of mathematics, network motifs are represented as connected subgraphs with a given number of vertices. There are 3-motifs (subgraphs with three connected vertices), 4-motifs (subgraphs with four connected vertices), and network motifs of a higher order [2–4]. There are 13 different (pairwise non-isomorphic) directed 3-motifs, and 2 undirected 3-motifs. The number of different directed 4-motifs is 198, the number of undirected 4-motifs is 6. Often, only statistically significant network motifs (that are found in the graph much more often than in randomized graphs) are called only motifs. We are splitting up these notions.

For example, protein-protein interaction networks and gene networks have significant motif, which is so-called BI-FAN motif (figure 1*a*). The reason for this is the structure of the interactions of receptors and substrates, kinase cascades and other types of protein-protein interactions [5]. A significant 3-motif for gene networks is a FEED-FORWARD LOOP motif, in which the target gene 3 is regulated by protein products of gene 1 and gene 2, with the protein product of gene 2 also regulated by the protein product of gene 1 (figure 1*b*). In food chains, the known significant 4-motif is a BI-PARALLEL motif, when the food base of organisms 2 and 3 is an organism 4, while both the organism 2 and organism 3 are a food base for an organism 1 (figure 1*c*). Significant network motifs [1] in digital circuits of fractional multipliers from the ISCAS'89 are cycles through three and four vertices (figures 1 *d*, 1 *e*).

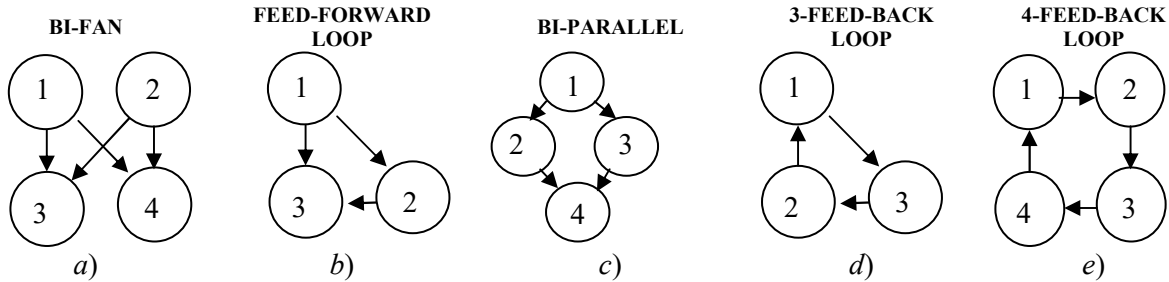

**Figure 1.** Significant 3- and 4-motifs

However, calculating the frequencies of network motifs in graphs is a complicated problem. Thus, in the well-known works [4–9], an acceptable time expenditure of exact calculating the frequencies of motifs was achieved only for graphs containing no more than twenty thousands of vertices, while many biological and social large networks contain millions of nodes and connections. Exact calculating the frequencies of network motifs in graphs of such large networks become unrealizable practically. Indeed, as it can be seen in table 1, the calculation of frequencies of 3- and 4- motifs by R environment (igraph package) and by the Fanmod program takes much time (calculations that could not be completed in more than 150 hours) are marked with a dash. The programs implement one of the fastest method for motifs calculating [6, 10]. We used the HP Z1 G2 Workstation with a 4-core Intel Xeon E3-1245 3.3 GHz CPU with Hyper-Threading technology and 8 GB DDR3-1333 RAM as a test bench. The molecular networks are available on the NDEx (Network Data Exchange) database. Structures of social networks are provided by the Leskovec's scientific team in Stanford University (http://snap.stanford.edu/data/).

**Table 1.** Time (in seconds) of calculation of frequencies for 3- and 4-motifs using R-Studio and Fanmod program

| Network | Nodes | Time of calculation | | | |
| --- | --- | --- | --- | --- | --- |
| | | Igraph | | Fanmod | |
| | | 3-motifs | 4-motifs | 3-motifs | 4-motifs |
| Protein-protein interactions network (Homo Sapience) | 22036 | 35 | 18824 | 381.026 | 182161 |
| Gene network GenReg | 16859 | 44 | 15163 | 513.986 | 188820 |
| Molecular network PathwayCommons | 19987 | 300 | 388056 | 323.9 | – |
| Fragment of social network (Google Plus) | 107614 | 39442 | – | – | – |
| Fragment of social network (Twitter) | 81306 | 76 | 35217 | 1075.69 | – |

Since the calculation of the frequency of motifs in large networks by exact methods requires too much time, special attention should pay to Monte Carlo methods to solve this problem.

## 2. Monte Carlo methods
The following three Monte-Carlo methods are proposed, which are used to calculate the frequency of motifs.
   1) The method of random motif sampling, called the method of ESA (Efficient Sampling Algorithm) [3]. In this method, a random motif sampling starts with endpoint vertices of randomly chosen edge, then the vertex set is iteratively extended by neighboring vertices until a desired size *k* is obtained. The motif having this set of vertices is detected. But the ESA method provides biased estimates of frequency of motifs.

2) The Rand-ESU method enumerates motifs sequentially, but some parts of the motifs could be cut off. This method uses ~~the~~ a so-called ESU-tree (the branches of this rooted tree on a given tree level *m* correspond to all motifs with *m* vertices). Some part of the ESU-tree branches is cut off so that you can finally get an unbiased estimate of frequencies of network motifs. The most efficiently calculation is an independent random cut off of tree branches on the last tree level.

3) Earlier, we proposed [11, 12] a method for calculating the frequencies of network motifs, which we call the method of frame sampling (MFS). And now it considers the ability to calculate directed 4-motifs, we developed a method for obtaining effective (minimized variance) estimates of frequencies of 4-motifs; also we unified the representation of motifs using canonical numbering of graphs. As it will be shown below, the new MFS (hereinafter referred to as the MFS) allows us counting the number of motifs with a given accuracy and does it faster than the known implementations of the Rand-ESU method.

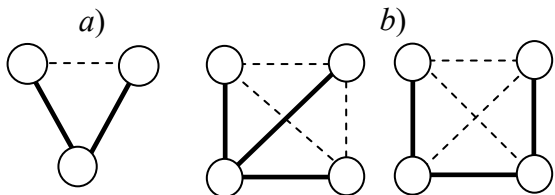

**Figure 2.** Bold lines are frames with three (*a*) and four (*b*) vertices. Dotted lines are edges that can be found 4-motive using the frame

The MFS [11] is an implementation of Monte Carlo method, in which a sample space is defined as a set of frame instances contained in the graph and which are sampling with equal probability. The number of different frames (and therefore the number of different sample spaces and various statistical experiments, which are used to calculate *k*-motifs) is equal to the number of different spanning trees with *k* vertices. There is the only spanning tree with three vertices (figure 2*a*), two spanning trees with four vertices, and three ones with five vertices. Figure 2*b* shows frames with four vertices: on the left it is a 'branching' frame, on the right it is a 'chain' frame.

The basis of the MFS is an equiprobable sample of the frame instances in the graph. The frame instance determines the motif which includes the frame instance. Since the total number of frame instances in a graph is known exactly (it is easily calculated), one can get unbiased estimate of frequency of motif instances in the graph.

An equiprobable sample of a frame with 3 vertices—forks (or paths of length 2, see Figure 2 a) is done in two steps. In the first step, a random vertex selection is performed. A vertex *i* is selected with probability $p_i$ proportional to the number of forks whose centers coincide with the vertex, $p_i = \binom{k_i}{2} N_V^{-1}$, where $k_i$ is a degree of the vertex *i*, $N_V$ – number of forks in the graph. At the second step, a pair of edges incident to the vertex *i* are chosen equally and a fork is formed. The probability of choosing the fork in the second step is $p' = \binom{k_i}{2}^{-1}$. The probability of choosing this fork is $p_i p' = 1/N_V$. An equiprobable sample of "branches" with 4 vertices (or trident, see figure 2*b*) from all tridents is similar, only the binomial coefficient $\binom{k_i}{2}$ is replaced by $\binom{k_i}{3}$.

For equiprobable sample of 'chains' (or paths of length 3, see figure 2*b*), in the first step, the edge of the graph is randomly selected. The probability to select the edge (*i*, *j*) is proportional to the product $(k_i - 1)(k_j - 1)$, i.e. the number of paths of length 3 in the middle of which there is a given edge. In the second step, an equiprobable selection of one of these paths is performed. As a result, in two steps, one of the sets of all chains in the graph is chosen equally. Sometimes the chosen path of length 3 is a closed chain (degenerate frame, i.e. a triangle). In this case, zero 4-motifs is detected on the chain, so the motif counters do not change, but the experiment counter (the number of chains scanned) is incremented by one.

If a frame or a set of frames (as in the case of 4-motifs) let us to detect all motifs with a given number *k* of vertices, then statistical experiments with an equally probable sample of such frames allow us to calculate the frequencies of all possible *k*-motifs.

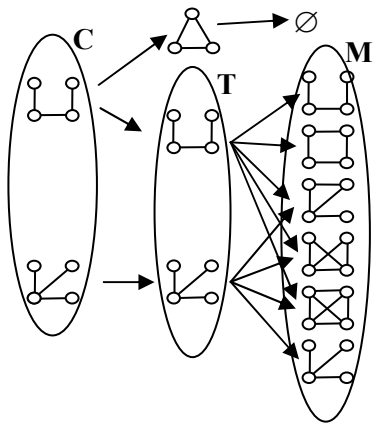

**Figure 3.** Surjection **C** →**T** →**M**, where **C** is a set of nondegenerate frames with four vertices, **T** is a set of spanning trees with four vertices, **M** is a set of 4-motifs. «Triangle» is degenerate frame 'chains'

For example, there is a bijection between the set C of nondegenerate frames with 4 vertices and the set T of their equivalent spanning trees (figure 3) and, accordingly, the map of the set of frames C to the set M of 4-motifs containing these frames is a surjection (figure 3).

Therefore, two sample spaces, each of which is defined on the set of instances of the corresponding frame in **C**, may contain all 6 motifs instances.

It should also be noted that if a certain motif is detected using *koef* different frame instances, then the estimation of the number of the motifs in a graph, obtained from a sample of frame instances, should be taken with the corresponding lowering coefficients 1/*koef*. Thus, the 4-motif 'clique' is detected to be six different non-degenerate chains (see figure 3), therefore, if on such randomly selected chains the 'clique' was encountered *s* times, then the number of different detected instances of the square is taken to be *s*/*koef*.

## 3. Mixed MFS

More than one statistical experiment is required to calculate the frequencies of network motifs with four or more vertices (to cover all network motifs). For example, the calculation of the frequencies of 4-motifs requires two different statistical experiments according to different spanning trees with 4 vertices (see figure 3).

When performing a statistical experiment *A* with a sample of chains (figure 4), the frequencies of five different motifs can be obtained. When performing experiment *B* with a sample of branches, the frequencies of four different motifs can be obtained. If motifs are found only in one of experiments, then estimates of frequency of motifs in the graph are computed in an obvious way. If the same motif is found in both experiments, then two different estimates of the frequency of the motif are obtained.

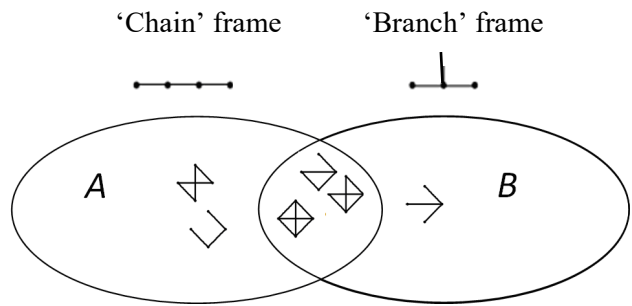

**Figure 4.** A diagram of the coverage of two different 4-motifs by frames: *A* is a set of motifs containing a 'chain' frame, *B* is a set of motifs containing a 'branch' frame

Let us consider this situation in details. Suppose that as a result of $N_A^+$ samples in experiment *A*, a network motif was found $C_A$ times. As a result of $N_B^+$ samples in experiment *B*, the same motif was discovered $C_B$ times. The relative frequencies of this motif in two experiments are equal $q_A = \frac{C_A}{N_A^+}$ and $q_B = \frac{C_B}{N_B^+}$, respectively. First network motif count estimate in a graph can be obtained from the results of experiment $A$ as $n_A = q_A N_A / \text{koef}_A = \dfrac{C_A}{N_A^+} N_A / \text{koef}_A$, where $N_A$ is the exact number of all frame A instances in the graph, the other estimate can be obtained as $n_B = q_B N_B / \text{koef}_B = \dfrac{C_B}{N_B^+} N_B / \text{koef}_B$, where $N_B$ is the exact number of all frame B instances. The 1/koefA and 1/koefB values are some correction lowering coefficients for the motif under study, these coefficients take into account how match frame instances can be contains of a motif. The mixed estimate $n$ of the number of the motif in the graph will be constructed as a linear combination of estimates

$$n = n_A + \lambda(n_B - n_A), \text{ где } 0 < \lambda < 1. \tag{1}$$

The numbers $C_A$ and $C_B$ of detections of the motif in $A$ and $B$ experiments determine the probabilities of detections of the motif in these experiments: $P_A \approx q_A = \dfrac{C_A}{N_A^+}$ and $P_B \approx q_A = \dfrac{C_B}{N_B^+}$. The following expressions can be obtained for the variances $D(n_A)$ and $D(n_B)$ of partial estimates:

$$D(n_A) \approx D\left(\dfrac{C_A}{N_A^+} N_A / \text{koef}_A\right) = \dfrac{N_A^2}{\text{koef}_A^2 (N_A^+)^2} D(C_A) = \dfrac{N_A^2}{\text{koef}_A^2 (N_A^+)^2} \sum_{i=1}^{N_A^+} P_A (1 - P_A) =$$

$$= \dfrac{N_A^2}{\text{koef}_A^2 (N_A^+)^2} N_A^+ \dfrac{C_A}{N_A^+}\left(1 - \dfrac{C_A}{N_A^+}\right) = \dfrac{N_A^2}{\text{koef}_A^2 (N_A^+)^2} C_A \left(1 - \dfrac{C_A}{N_A^+}\right),$$

and $D(n_B) = \dfrac{N_B^2}{\text{koef}_B^2 (N_B^+)^2} C_B \left(1 - \dfrac{C_B}{N_B^+}\right)$.

The optimal value of the parameter $\lambda$ is determined from the condition of minimizing the coefficient $v$ of the variation of the estimate

$$v^2(\lambda) = \dfrac{D(n(\lambda))}{M^2(n(\lambda))} \to \min, \ 0 < \lambda < 1. \tag{2}$$

Find the optimal $\lambda$ by differentiating $v^2$ by $\lambda$ and equating the resulting expression to zero

$$[v^2(\lambda)]' = \left[\dfrac{D(n(\lambda))}{M^2(n(\lambda))}\right]' = \left[\dfrac{D((1-\lambda)n_A) + D(\lambda n_B)}{M^2((1-\lambda)n_A + \lambda n_B)}\right]' = \left[\dfrac{(1-\lambda)^2 D(n_A) + \lambda^2 D(n_B)}{((1-\lambda)n_A + \lambda n_B)^2}\right]' =$$

$$= \dfrac{2D(n_B)\lambda - 2D(n_A)(1-\lambda)}{(n_A(1-\lambda) + n_B \lambda)^2} - \dfrac{2(n_B - n_A)(D(n_A)(1-\lambda)^2 + D(n_B)\lambda^2)}{(n_A(1-\lambda) + n_B \lambda)^3} =$$

$$= \dfrac{[2D(n_B)\lambda + 2D(n_A)(1-\lambda)](n_A(1-\lambda) + n_B \lambda) - 2(n_B - n_A)(D(n_A)(1-\lambda)^2 + D(n_B)\lambda^2)}{(n_A(1-\lambda) + n_B \lambda)^3} = 0. \tag{3}$$

Dividing equality (3) by a common nonzero factor $2 \cdot /(n_A(1-\lambda) + n_B \lambda)^3$, we get the equation

$$n_A D(n_B)\lambda(1-\lambda) + n_B D(n_B)\lambda^2 - n_A D(n_A)(1-\lambda)^2 - n_B D(n_A)\lambda(1-\lambda) -$$
$$- n_B D(n_A)(1-\lambda)^2 - n_B D(n_B)\lambda^2 + n_A D(n_A)(1-\lambda)^2 + n_A D(n_B)\lambda^2 = 0,$$

after simplifying we obtain

$$n_A D(n_B)\lambda - n_B D(n_A) + n_B D(n_A)\lambda = 0.$$

From here we find the optimal $\lambda$ in the form

$$\lambda = \frac{n_A D(n_A)}{n_A D(n_B) + n_B D(n_A)}. \qquad (4)$$

Replacing $n_A$, $D(n_A)$, $n_B$, $D(n_B)$ with their expressions in terms of parameters calculated in both experiments, we obtain the following estimate for the optimal $\lambda$:

$$\lambda \approx \frac{\dfrac{C_A}{N_A^+} N_A \dfrac{N_A^2}{\text{koef}_A^2 (N_A^+)^2} C_A \left(1 - \dfrac{C_A}{N_A^+}\right)}{\dfrac{C_A}{N_A^+} N_A \dfrac{N_B^2}{\text{koef}_B^2 (N_B^+)^2} C_B \left(1 - \dfrac{C_B}{N_B^+}\right) + \dfrac{C_B}{N_B^+} N_B \dfrac{N_A^2}{\text{koef}_A^2 (N_A^+)^2} C_A \left(1 - \dfrac{C_A}{N_A^+}\right)}.$$

Using the value $\lambda$ calculated by this formula in (1), we calculate the optimal estimate of the number $n$ of motif instances in the graph.

**4. Using of recalculated canonical numbers of graphs**
We propose to the pre-calculated numbers of the graphs in the given canonical form (figure 5) while implementing the MFS. In this case, we will store all possible correspondences of graphs of a given size to the canonical numbers of motifs in the *arrcode* arrays. The sizes of these arrays for motifs with three and four vertices are presented in table 2.

**Table 2.** *Arrcode* **array sizes for storing graph numbers**

| Number of verticies | Type of motifs | Size of arrcode arrays |
|---|---|---|
| 3-motif | Directed | $2^6$ |
|  | Undirected | $2^3$ |
| 4-motif | Undirected | $2^6$ |
|  | Directed | 4096 |

The arrcode array contains the canonical numbers of motifs (we used BLISS codes). The size of the arrcode array is given by a number of possible combinations of edges (motifs are encoded by edges). So, for undirected 3-motifs arrcode = {0,1,1,2,1,2,2,3}. The first isomorphic class of arrcode (numbered 0) corresponds to set of isolated vertices; the last class (numbered 3) is a complete graph. The number of isomorphism classes for directed graphs with three vertices is 16 (between 0 and 15), for undirected only four. In both cases subgraphs containing at least one isolated vertex are also taken into account. There are 218 (directed graph) and 11 (undirected graph) isomorphism classes in graphs containing 4 vertices.

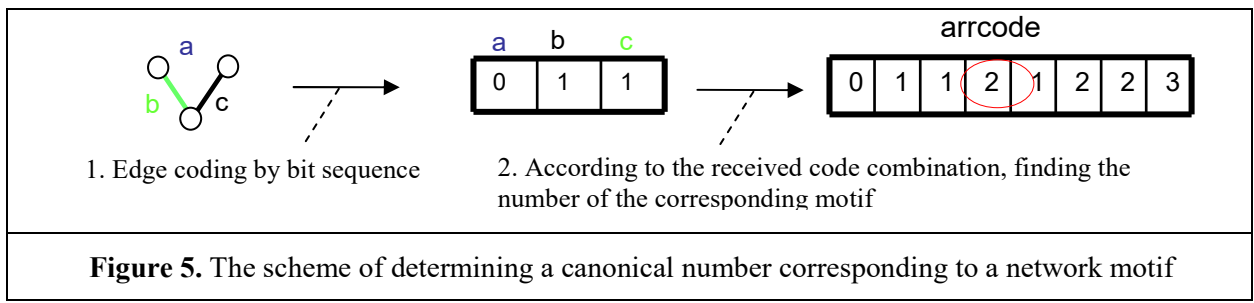

1. Edge coding by bit sequence
2. According to the received code combination, finding the number of the corresponding motif

**Figure 5.** The scheme of determining a canonical number corresponding to a network motif

**5. Calculation of directed graphs**
The problem of directed motif calculation can be solved technically. A random frame sample should be performed without arc orientation, then it is determined to which network motif was detected on given vertices triple (figure 6) or four vertices (figure 7).

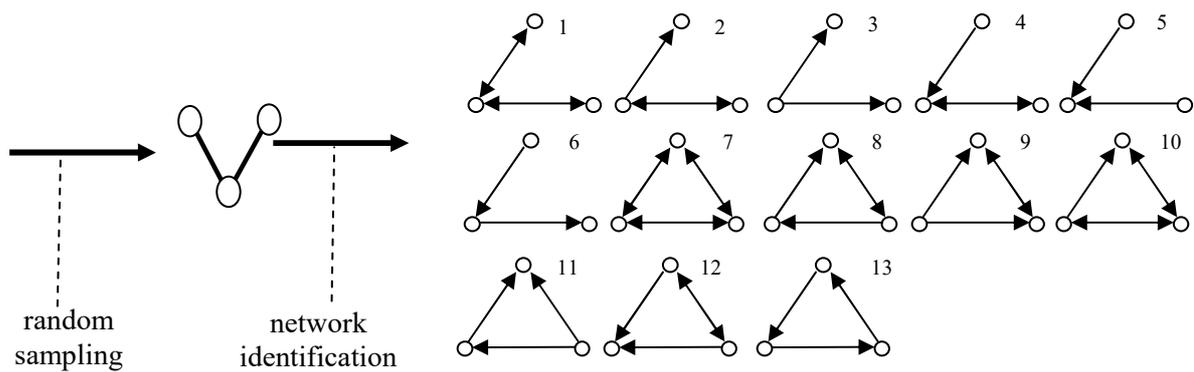

**Figure 6.** Expansion of the sample space by taking into account the arc directions (3-motifs)

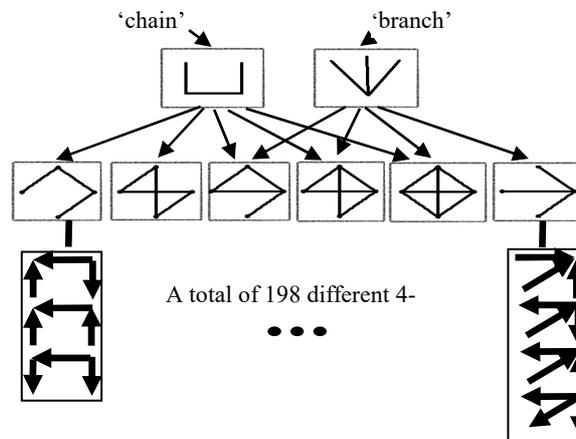

A total of 198 different 4-

**Figure 7.** Expansion of the sample space by taking into account the arc directions (4-motifs)

## 6. Experimental results

A comparison of the Rand-ESU and the MFS with the same accuracy is performed. As can be seen in table 3 the MFS significantly speeds up the calculation with compare to the Rand-ESU, at the same time not less accurate calculations are achieved. So, 100 thousand motifs were sampled by Rand-ESU method and 200 thousand frames were sampled by MSF.

**Table 3.** Comparison of the calculation time for the 4-motifs by MFS and by the Rand-ESU (implemented in igraph and Fanmod programs)

| Network | Time of calculation, seconds | | |
|---|---|---|---|
| | Rand-ESU igraph | Rand-ESU Fanmod | MFS |
| Protein-protein interactions network (Homo Sapience) | 338 | 451 | 67 |
| Molecular network PathwayCommons | 3337 | 315 | 48 |
| Gene network GenReg | 677 | 756 | 15 |
| Fragment of social network (Twitter) | 1354 | 1255 | 36 |
| Fragment of social network (Google Plus) | 719333 | 386828 | 2642 |

## 7. Conclusion

We have significantly improved the method [11] of sampling frame (MSF).

Firstly, a scheme has been proposed for obtaining of effective (in terms of minimizing the error of calculation) 4-motifs frequencies estimates using statistical experiments with different frames.

Secondly, the problem of analyzing directed graphs is solved. The developed MSF has advantages compared with the popular Rand-ESU method. So, MFS is considerably less demanding in memory and less sensitive to the size of the network under study.

**Acknowledgments**

The reported study was funded by Russian Foundation for Basic Research, according to the research project No 16-31-60023 mol_a_dk.

**References**

[1] Milo R, Shen-Orr S, Itzkovitz S, Kashtan N, Chklovskii D, Alon U 2002 Network motifs: simple building blocks of complex networks *Science* **298** 824–7
[2] Shen-Orr S S, Milo R, Mangan S, Alon U 2002 Network motifs in the transcriptional regulation network of Escherichia coli *Nat. Genet.* **31** 64–8
[3] Itzhack R, Mogilevski Y, LouzounY 2007 An optimal algorithm for counting network motifs *Physica A: Statistical Mechanics and its Applications* **381** 482–90
[4] Schreiber F, Schwöbbermeyer H 2005 Frequency concepts and pattern detection for the analysis of motifs in networks *Transactions on Computational Systems Biology III. Lecture Notes in Computer Science* **3737** 89–104
[5] Ma'ayan A, et al 2005 Formation of Regulatory Patterns During Signal Propagation in a Mammalian Cellular Network *Science* **309** 1078–83
[6] Wernicke S, Rasche F 2006 FANMOD: a tool for fast network motif detection *Bioinformatics* **22** 1152–3
[7] Chen J, Hsu W, Li Lee M, et al 2006 NeMoFinder: dissecting genome-wide protein-protein interactions with meso-scale network motifs T*he 12th ACM SIGKDD international conference on Knowledge discovery and data mining* Philadelphia Pennsylvania USA 106–15
[8] Kashani Z R, Ahrabian H, Elahi E, et al 200 Kavosh: a new algorithm for finding network motifs *BMC Bioinformatics* **318** 1–12
[9] Meira Luis A, Vinicius A, Maximo R, et al 2014 Ace-Motif: accelerated network motif detection IEEE/*ACM Computational Biology and Bioinformatics* **11** 853–62
[10] Smoly I Y, et al 2017 MotifNet: A web-server for network motif analysis *Bioinformatics* **33** 1907–9
[11] Yudin E B, Zadorozhnyi V N 2015 Statistical approach to calculation of number of network motifs *International Siberian Conference on Control and Communications* 7147296–9
[12] Yudin E B, Yudina M N 2017 Calculation of number of motifs on three nodes using random sampling of frames in networks with directed links *Siberian Symposium on Data Science and Engineering (SSDSE)* 8071957–60